\documentstyle[12pt,aaspp4]{article}

\newcommand{\kms}{$\,\mbox{km}\,\mbox{s}^{-1}$}
\newcommand{\ea}{\emph{et al.\ }}

\slugcomment{accepted for publication in the ApJ Letters}

\lefthead{Walter \ea}
\righthead{X--ray Emission from an Expanding Supergiant Shell in IC\,2574}

\begin{document}

\title{X--ray Emission from an Expanding Supergiant Shell in IC\,2574}

\author{Fabian Walter}
\affil{Radioastronomisches Institut der
Universit\"{a}t, Auf dem H\"{u}gel 71, D--53121 Bonn, Germany
\\Electronic mail: walter@astro.uni-bonn.de} 
 
\author{J\"urgen Kerp}
\affil{Radioastronomisches Institut der
Universit\"{a}t, Auf dem H\"{u}gel 71, D--53121 Bonn, Germany
\\Electronic mail: jkerp@astro.uni-bonn.de} 

\author{Neb Duric}
\affil{Department of Physics and Astronomy, University of New Mexico,
Albuquerque, NM 87131\\Electronic mail: duric@deneb.phys.unm.edu} 
 
\author{Elias Brinks}
\affil{Departamento de Astronom\'{\i}a, Apartado Postal 144,
Guanajuato, Gto.  36000, M\'exico\\National Radio Astronomy Observatory, 
P.O. Box O, Socorro, NM 87801 \\Electronic mail: ebrinks@astro.ugto.mx} 

\and

\author{Uli Klein}
\affil{Radioastronomisches Institut der
Universit\"{a}t, Auf dem H\"{u}gel 71, D--53121 Bonn, Germany
\\Electronic mail: uklein@astro.uni-bonn.de} 

%
%

\begin{abstract}

We present a multi--wavelength study of a supergiant shell within the
violent interstellar medium of the nearby dwarf galaxy IC\,2574, 
a member of the M81 group of galaxies.  Neutral hydrogen (\ion{H}{1}) 
observations obtained with the Very Large Array (VLA) reveal a prominent 
expanding supergiant \ion{H}{1} shell in the northeast quadrant of 
IC\,2574 which is thought to be produced by the combined effects of 
stellar winds and supernova explosions. It measures roughly $1000 
\times 500$ pc in size and is expanding at about 25~\kms. The \ion{H}{1} 
data suggest an age of about $1.4 \times 10^6$ yrs; the energy
input must have been of order $(2.6\,\pm\,1) \times 10^{53}$ ergs. 
Massive star forming regions, as traced by H$\alpha$ emission, are 
situated predominantly on the rim of this \ion{H}{1} shell. This supports 
the view that the accumulated \ion{H}{1} on the rim has reached 
densities which are high enough for secondary star formation to commence.
 VLA radio continuum observations at $\lambda$\,6~cm show that these 
star-forming regions are the main sources of radio continuum emission in 
this galaxy. This emission is mainly thermal in origin. Soft X--ray 
emission from within the \ion{H}{1} hole is detected by a pointed 
ROSAT PSPC observation. The emission is resolved, coinciding in  
size and orientation with the \ion{H}{1} shell. These spatial properties 
suggest that the emission is generated by an X--ray emitting plasma 
located within the \ion{H}{1} shell although a contribution from 
X--ray binaries cannot be completely ruled out. The X--ray luminosity 
within the 0.11--2.4\,keV energy range is  $\rm L_X\,=\,(1.6\,\pm\,0.5)
\times10^{38}$\,ergs\,s$^{-1}$. The X--ray data are compatible with 
emission coming from a Raymond \& Smith (\markcite{RAY77}1977) plasma at 
a temperature of about log($T\rm [K])\,=\,6.8$ and a density of 
0.03\,cm$^{-3}$. The energy content of the coronal gas corresponds 
to $(4\,\pm\, 2)\times 10^{53}\,$ergs, or broadly in agreement with the 
energy input derived on the basis of the \ion{H}{1} observations.

\end{abstract}

%
%


%
%

\section{Introduction}

The interstellar medium (ISM) of galaxies is shaped by the aftermath of 
recent, massive star formation (SF). Huge cavities filled  with coronal 
gas are believed to be blown in the ambient medium through 
strong stellar winds of the most massive stars as well as through 
subsequent Type II supernova (SN) explosions of stars more massive than
$\approx$8~M$_\odot$ (see Tenorio--Tagle \& Bodenheimer \markcite{TEN88}1988 
for a review). The structures that form during such processes are often 
referred to as bubbles (diameter $\sim$10~pc), superbubbles 
($\sim$200~pc) 
or supergiant shells ($\sim$1~kpc). A recent development is the finding 
that supergiant shells reach much larger sizes in dwarf galaxies than 
in more massive spiral galaxies (Puche \ea \markcite{PUC92}1992, 
Walter \&~Brinks 1998 \markcite{WAL98} --- hereafter referred to as 
WB98). This is attributed to dwarfs having a lower gravitational 
potential, 
which allows shells to grow more easily, and to the absence of
differential
rotation and spiral density waves, which prevents them from being
destroyed prematurely. Dwarf galaxies are therefore ideally suited to 
study the origin and evolution of supergiant shells.

Cox \& Smith (\markcite{COX74}1974) not only realized that the energy
input of SNe creates cavities filled with hot gas, but that the
cooling time is relatively long ($10^6$ to $10^9$ years, depending on
density and temperature of the matter within), allowing the creation
of a tunneling network (the swiss cheese model). Weaver \ea 
(\markcite{WEA77}1977) discussed a model in which fast stellar winds 
are adiabatically shocked to temperatures of order $10^6$--$10^7$~K. 
These winds then act like a piston, driving the expansion of the outer 
shell of swept-up ambient material. The basic picture still stands 
today, although several refinements have been proposed, e.g., by Chu
\& Mac~Low (\markcite{CHU90}1990) who calculate the expected X--ray
emission from superbubbles. 

The ISM that is swept up accumulates in the form of an expanding shell and
forms the rim of a superbubble or supergiant shell. The resulting shocks 
on their rims may induce new generations of SF, or SF regions may be 
created simply via gravitational fragmentation (Elmegreen 
\markcite{ELM94}1994, for an early discussion see Mueller \& Arnett 
\markcite{MUL76}1976). 

Direct observational evidence  for the theoretical scenarios mentioned
above is still scarce. Very few, if any, supergiant shells have been
detected to date through the emission of the coronal gas (see Bomans
\ea \markcite{BOM96}1996 for a review). In fact, it is the neutral and
ionized gas which has proven to be the best tool to trace the outlines
of superbubbles. The best direct evidence has come from  
ROSAT PSPC observations, giving information about the plasma 
properties of the interior of some superbubbles. Examples are the
supergiant 
shell LMC\,4 (Bomans, Dennerl \& K\"urster \markcite{BOM94}1994), the 
superbubbles N\,44 (Kim \ea \markcite{KIM98}1998) and N\,11 
(Mac~Low \ea \markcite{MAC98}1998), all three situated in the Large 
Magellanic Cloud (LMC), the supergiant shell SGS\,2 in NGC\,4449 
(Bomans, Chu \& Hopp \markcite{BOM97}1997) and the possible supershell 
near Holmberg\,IX (Miller \markcite{MIL95}1995). In the following we 
will present the probable detection of coronal gas filling a supergiant 
shell within the nearby (3.2~Mpc) dwarf galaxy IC\,2574.

%
%

\section{Observational Evidence for a Hot--Gas filled expanding 
Supergiant Shell in IC\,2574}

\subsection{\ion{H}{1} observations}

The neutral interstellar medium of IC\,2574 has been extensively 
studied by WB98 using \ion{H}{1} observations obtained with the 
NRAO\footnote{The National Radio Astronomy Observatory (NRAO) is 
operated by Associated Universities, Inc., under cooperative
 agreement with the National Science Foundation.} Very Large Array. 
An \ion{H}{1} surface brightness map of IC\,2574 is shown in Fig.~1\,a. 
WB98 find a total of 48 \ion{H}{1} holes in IC\,2574 most of which are 
expanding. The most prominent of these holes (No.\ 35 in their study) is 
the subject of this letter. An \ion{H}{1} map of the region around this 
hole is shown as a blowup in Fig.~2\,a. The size of the hole is 
$\approx$1000~pc $\times$ 500~pc (which corresponds to an effective radius 
$\rm r_{\rm eff}\approx350$~pc) and its radial expansion velocity is 
$\approx$25~\kms. The dynamical age is therefore about $1.4\times 10^7$ 
years. WB98 show that the 1$\sigma$--scale height of the \ion{H}{1} 
disk is approximately 350~pc,
or of the same order as the radius of the hole. Hence, the hole is
still contained within the disk (no material lost due to blow--out of
the gas into the halo). Given the derived scaleheight and ambient
average \ion{H}{1} column density of $N_{\rm HI}\approx 10^{21}\,\rm
cm^{-2}$, the ambient \ion{H}{1} volume density before the creation of
the hole is estimated at $\rm n_{HI}=0.5~cm^{-3}$. The \ion{H}{1} mass
that was present before the evacuation of the \ion{H}{1} hole which
now must be largely accumulated on the rim is therefore about $2
\times10^6$~M$_{\odot}$. Using these numbers and correcting the
\ion{H}{1} mass for the contribution of primordial helium, we estimate
the kinetic energy of the expanding shell to be $\approx 1.7\times
10^{52}$~ergs.

No signature of an infalling high--velocity cloud (HVC) is found 
in the \ion{H}{1} data and WB98 propose an internal origin for the 
creation of the hole. Using the numerical models of Chevalier 
(\markcite{CHE74}1974), who calulates the late phase of supernova
remnants in which the shell accretes matter from both the interstellar
medium and the hot 
interior, an energy input of $(2.6\,\pm\,1)\times 
10^{53}$~ergs by a star forming region is needed to create this hole. 
Note that the kinetic energy of the hole is of order 10\% of this 
value, which is about half the value originally predicted by Weaver 
\ea \markcite{WEA77}(1977). This energy is equivalent to about 100 
Type II SNe and the strong stellar winds of their progenitors; a 
major SF event.

\subsection{H$\alpha$ observations}

In order to trace regions of current SF activity, H$\alpha$ maps were 
obtained by WB98. A greyscale representation of their H$\alpha$ map is 
given in Fig.~1\,b. Although the global SF activity in IC\,2574 is 
rather low,  a ring of \ion{H}{2} regions is prominent in the northeast 
quadrant of the galaxy. A blowup of the H$\alpha$  emission around the 
same region as in Fig.~2\,a is given in Fig.~2\,b. 
Note that virtually all H$\alpha$ emission in that region is situated 
on the rim of the \ion{H}{1} shell (as first noticed by Martimbeau, 
Carignan \& Roy \markcite{MAR94}1994). Three Wolf--Rayet stars have been 
detected in the most luminous \ion{H}{2} region on the northern rim 
by  Drissen, Roy \& Moffat (\markcite{DRI93}1993). IUE spectra through 
the same region show the presence of P~Cygni profiles indicating the 
presence of strong stellar winds (Rosa, Joubert \& Benvenuti 
\markcite{ROS84}1984). Using the calibration by Miller \& Hodge 
(\markcite{MIL94}1994), the total H$\alpha$ flux on the rim is 
$2.3\times 10^{-12}$~erg~cm$^{-2}$\,s$^{-1}$ which corresponds to a 
luminosity of $L(\rm H\alpha)=7\times 10^5$ L$_{\odot}$ 
(L$_{\odot}=3.85\times 10^{33}$~erg~s$^{-1}$) or an energy equivalent 
of about 50 stars of spectral type O5 (WB98). This is almost one third 
of the H$\alpha$ luminosity of 30~Dor, which is believed to be the 
most active SF region in the Local Group (Chu \& Kennicutt 
\markcite{CHU94}1994). The average volume densities derived on the 
rim of the shell are 3~cm$^{-3}$ for the neutral \ion{H}{1} 
(southern part) and 2~cm$^{-3}$ for the ionized hydrogen in the 
northern part. This suggests that most of the neutral gas component 
in the northern part of the hole has been ionized by the prominent SF 
regions.

\subsection{Radio Continuum observations}

Radio continuum maps of IC\,2574 were obtained at $\lambda\,$6 cm
using the VLA D--array. A map of the  $\lambda\,$6--cm emission is 
presented in Fig.~2\,d.  We detect four radio sources in the region of 
the supergiant shell; their positions and flux densitities (in parentheses)
are B102459.7+684320 (S\,=\,0.9\,mJy), B102503.1+684348
(S\,=\,1.1\,mJy), B102507.5+684325 (S\,=\,0.8\,mJy) and B102507.7+684346
(S\,=\,0.8\,mJy). The flux densities are accurate to within 20\%. Three 
of the 
four sources coincide with regions of intense H$\alpha$ emission on the
rim of the shell. These sources have Radio/H$\alpha$ flux ratios that are
consistent with the radio emission being purely thermal (Caplan \& 
Deharveng 
\markcite{CAP86}1986). This result confirms that the H$\alpha$ emission 
traces photo--ionized gas and that these sources are indeed \ion{H}{2} 
regions and not shock--ionized regions due to the expansion of the hole. 

The source furthest to the southwest, B102459.7+684320, is clearly 
nonthermal (see Fig.~2\,d) and its radio emission is well in excess 
of what one would expect from an \ion{H}{2} region on the basis of the 
rather weak H$\alpha$ emission. It most likely is a
supernova remnant given its nonthermal nature, its appearance as a
point source and a flux density which is consistent with the brighter 
SNRs found in nearby galaxies (e.g., Duric \ea \markcite{DUR95}1995; 
Lacey \ea \markcite{LAC97}1997).

\subsection{X--ray observations}

In order to examine the X--ray properties of IC\,2574, we have analysed
a ROSAT PSPC observation towards IC\,2574 (US600145P,
total integration time: 7.3\,ks). The PSPC point spread function at 
${\overline E}\,=\,1.23$\,keV is about $28''$ (FWHM) across the
area 
of interest. The pointed ROSAT PSPC observation reveals a significant
 X--ray feature which is coincident with the very center of the \ion{H}{1} 
hole.  A contour map of the X--ray emission is overlaid on the
H$\alpha$--emission in Figs.~1\,b (whole galaxy) and 2\,b (blowup). An
overlay with the \ion{H}{1} is presented in Fig.~2\,c.
This X--ray feature has an angular extent of 
about $45''\,\times\,30''$ (FWHM), with its
major axis aligned with the \ion{H}{1} hole. Note that the emission is
extended even if we account for the residual pointing uncertainties of
the ROSAT telescope of about $10''$ (Hasinger 1997, priv.\ comm.). In
addition, ROSAT's wobble direction during the observations was
perpendicular to the orientation of the major axis of the \ion{H}{1}
hole. Based on the ROSAT PSPC observations we therefore find evidence for 
extended X--ray emission originating from the interior of the 
\ion{H}{1} hole. However higher resolution observations will be 
needed to confirm this.

IC\,2574 is located at a high galactic latitude ($b\approx+43^{\circ}$), 
which means that the photoelectric absorption due to the Galactic 
interstellar medium is moderate. The Leiden/Dwingeloo \ion{H}{1} 
$\lambda$\,21-cm line survey (Hartmann \& Burton
\markcite{HAR97}1997) gives an \ion{H}{1} column density of 
$N_{\rm HI}\,=\,2\times10^{20}\,{\rm cm^{-2}}$ after separating the
Galactic diffuse \ion{H}{1} emission from that of IC\,2574.
Because of the apparently clear line of sight through the hole (see
Fig.~1\,a) we considered it justified to neglect any contribution to
the \ion{H}{1} column density within  IC\,2574.

To separate the source counts from that of the diffuse X--ray background,
we
performed an on/off analysis. This method also removes the contribution of 
the so-called non-cosmic X--ray background radiation (Snowden \ea 
\markcite{SNO94}1994). In order to obtain quantitative information 
about the spectrum of the
X--ray feature, we evaluated the standard ROSAT hardness ratios (HR1
and HR2), which range from -1 (soft) to 1 (hard). We took this
approach because the number of photons is too low for a standard
analysis of the X--ray spectrum. We divided the ROSAT energy band into
the three main energy bands, denoted as ROSAT C-band (0.25\,keV, PI
channels 11-41) the M-band (0.75\,keV, 52-90) and J-band (1.5\,keV,
91-201). The evalution of HR2 which is nearly unaffected by galactic
absorption yields a value of HR2\,=\,$\frac{(J-M)}{(J+M)}=0.2\pm 0.2$.
This leads to a first estimate of the temperature of log(T[K])\,$\approx$\,6.8.
Adopting this estimate we can correct the soft bands for
photoelectric absorption by the Galactic ISM (Snowden \ea
\markcite{SNO94}1994). This analysis yields
HR1\,=\,$\frac{(J+M-C)}{(J+M+C)}=0.2\pm 0.4$ and the following photon
counts for energies below and above 1 keV: $I(\rm E\,<\,1\,keV,
11-90)\,=\,106\,\pm\,39$\,cts and $I(\rm E\,>\,1\,keV,
91-201)\,=\,38\,\pm\,9$\,cts.  This simple approach indicates that the
X--ray feature emits most of its photons below
$E\,=\,1$\,keV. Combined with the fact that the source is elongated,
this suggests that a high mass, or even low mass X--ray binary is
unlikely to be the source of the detected emission. It should be
noted though that a possible contribution from X--ray binaries to the
observed X--ray emission cannot be ruled out completely at this point.

Using the hardness ratios and assuming a reduced metallicity of about
15\% solar (as derived in a few \ion{H}{2} regions in IC\,2574 by
Miller \& Hodge \markcite{MIL94}1994), we calculate a plasma
temperature range, according to Raymond \& Smith
(\markcite{RAY77}1977), of log($T_{\rm 15\%}$[K])\,=\,6.9\,--\,7.1.
If the X--ray plasma within the \ion{H}{1} hole contains more metals
the temperature will be lower (log($T_{\rm
solar}$[K])\,=\,6.4\,--\,6.8).  Note that a higher abundance is more
likely since the interior of the shell has recently been enriched by
the heavy elements ejected by the SNe that created the hole. As a
result the true temperature will lie somewhere in between; we adopt a
temperature of log($T\rm [K])\,=\,6.8\pm0.3$ for the remainder of this
paper.

For a plasma of log($T\rm [K])\,=\,6.8\pm 0.3$ and a reduced metallicity
we evaluate an emission measure of  $EM\,=\,(0.65\,\pm\,0.15)\,
{\rm cm^{-6}\,pc}$. Assuming an extent of the line of sight through the 
X--ray cavity of 700\,pc, we derive an electron density of 
$n_{\rm e}\,=\,(0.03\,\pm\,0.01)\,{\rm cm^{-3}}$. This value is similar to
the densities found in the supergiant shell SGS2 in NGC\,4449
(0.03$\rm\ cm^{-3}$) or in the LMC\,4 ($\approx 0.01\rm\ cm^{-3}$) (Bomans
\ea
\markcite{BOM97}1997). The total X--ray luminosity of the interior of
the hole is $\rm L_X(0.28-2.4\,keV)\,=\,(1.6\,\pm\,0.5)
\times10^{38}$\,ergs\,s$^{-1}$ which lies in between the values for
SGS\,2 in NGC\,4449 ($\approx 7 \times10^{38}$\,ergs\,s$^{-1}$) and LMC\,4
($\approx 0.15 \times10^{38}$\,ergs\,s$^{-1}$). This luminosity is about one
order of magnitude higher than predicted by theoretical models (see, e.g.,
the calculations by Chu \ea 1995), as is the case with most X--ray
luminosities
determined thus far (N11 being the only exception to date -- see 
the discussion by Mac~Low \ea \markcite{MAC98}1998).

%
%

\section{Discussion} 

If we assume that the supergiant shell is the result of a recent SF 
event, 
the picture which emerges from the observations discussed above is
the following: about $1.4\times10^7$~years ago (the dynamical age of the
\ion{H}{1} hole), a major SF event took place at the center of what 
today shows up as a prominent \ion{H}{1} hole. Since its creation
some $2.6\times 10^{53}$~ergs of energy have been deposited by the 
most massive stars, into the ambient ISM. About 10\% of this energy 
is now still present in the form of the kinetic energy of the expanding 
shell. Note that the least massive stars that go off as SN are most 
probably still present in the cavity since their lifetime ($\approx 5 
\times 10^7$\, years) is somewhat longer then the dynamical age of the 
hole. In other words, there is still some energy input going on today 
which might explain why the derived temperature for the interior 
of the \ion{H}{1} hole is at the upper end for supergiant shells. 
On the other hand, a larger than expected contribution from X--ray
binaries may also explain the high temperature estimate.
Adopting a sound speed of 100~\kms\ for the coronal gas filling the hole, 
we derive a time of only $\approx 3\times 10^6$~years before a soundwave 
actually reaches the rim of the shell. The gas has therefore 
had enough time to establish a relatively uniform distribution within 
the cavity. For a plasma temperature of log($T\rm [K])\,=\,6.8\pm0.3$ 
and an internal density of $(0.03\,\pm\,0.01)$\,cm$^{-3}$ we derive 
an internal pressure of $P=2\,n_{\rm e}\,T\approx (4\,\pm \,2) 
\times10^5$\,K\,cm$^{-3}$ (LMC\,4: P $= 2\times 10^4$\,K\,cm$^{-3}$). 
This value has to be compared to the much lower pressure of the ambient 
warm ionized medium ($P\approx 10^3-10^4$\,K\,cm$^{-3}$). This means 
that it is probably this hot gas which is still driving the expansion 
of the shell (see, e.g., Weaver \ea \markcite{WEA}1977). The shell, as 
it is sweeping up material, has already reached the point at which 
secondary star formation can take place, as evidenced by the \ion{H}{2} 
regions located along its perifery. The energy of the hot--gas interior 
can be approximated by  $\frac{3}{2}\,(2\,n_{\rm e})\,k\,T\,V$ and yields 
a value of $(4\,\pm\, 2)\times 10^{53}\,$ergs. The model by Weaver \ea 
(\markcite{WEA77}1977) predicts that about 50~\% of the total energy 
deposited by SNe goes into the thermal heating of the gas interior a 
superbubble. Our values indicate that, within the errors, indeed most 
of the energy deposited by the SN explosions has gone into the heating 
of the gas. Note that a small decrease in temperature and increase 
in the ambient density would bring these numbers in even closer 
agreement. 
In fact, it is the broad agreement of the observed numbers with those
derived from  theory which gives us added confidence that it is indeed 
hot 
diffuse gas that fills the cavity rather than a cluster of X--ray
binaries.
A definitive answer will have to await the better spectral and spatial 
resolution of future X--ray missions, such as AXAF and XMM, which are 
expected to shed further light on this unique object.

\acknowledgements

FW gratefully acknowledges the 'Deutsche Forschungsgemeinschaft (DFG)'
for the award of a stipendium in the Graduate School "The Magellanic
Clouds and other Dwarf Galaxies". EB acknowledges a grant from CONACyT
(0460P--E) and thanks the Graduate School for financially
supporting a visit to Bonn during which the foundations for this paper
were laid. We thank Dominik Bomans, You--Hua Chu, and Mordecai--Mark
Mac~Low for fruitful discussions during the Dwarf Galaxy workshop
organised by the Graduate School and an anonymous referee for valuable
comments which helped us to improve this paper.

\clearpage

\begin{figure}
\figcaption{Fig.\ 1\,a: \ion{H}{1} surface brightness map of IC\,2574 as 
presented in WB98. Note that the ISM of IC\,2574 is dominated by many 
hole--like structures. The dashed square indicates the \ion{H}{1} hole 
which is the focus of this letter. The dotted square shows the extent 
of the optical H$\alpha$--map. Fig.\ 1\,b: H$\alpha$--map of IC\,2574 
as presented in WB98 (greyscale). The contours (plotted at 15, 20, 30, 
50, 70, 90\% of the peak emission) represent the X--ray emission of 
IC\,2574 as observed with the ROSAT PSPC camera. The dashed square
again indicates the supergiant shell which is the focus of this letter.}
\end{figure}

\begin{figure}
\figcaption{Fig.\ 2\,a: Blowup of the \ion{H}{1} surface brightness map around the 
prominent \ion{H}{1} hole (the size of the beam is indicated in the
lower left). Fig.\ 2\,b: Blowup of the H$\alpha$--image with X--ray 
contours superimposed. The area shown is identical to that in Fig.\ 2\,a; 
the X--ray contour levels are the same as in Fig.\ 1\,b. Fig.\ 2\,c:
X--ray contours overlaid on the \ion{H}{1} hole. Fig.\ 2\,d: 
5$\lambda$\,6\,cm radio continuum emission of IC\,2574 (contours) 
overlaid on the H$\alpha$ map (greyscale). The beamsize is indicated in 
the lower left. Contours are plotted at 0.14, 0.22, 0.31, 0.39 and 0.48 
mJy~beam$^{-1}$.}
\end{figure}


\begin{references}

\reference{BOM94} Bomans, D.J., Dennerl, K., \& K\"urster, M. 1994, \aap, 
283, L21

\reference{BOM96} Bomans, D.J., Chu, Y.--H., Magnier, E.A., \& Points, S. 
1996, in R\"ontgenstrahlung 
from the Universe, ed.\ H.U.\ Zimmermann, J.E.\ Tr\"umper \& H.\ Yorke 
(Garching: Max--Planck Institut f\"ur Extraterrestrische Physik), p.\ 237 

\reference{BOM97} Bomans, D.J., Chu, Y.--H., \& Hopp, U. 1997, \aj, 113,
1678

\reference{CAP86} Caplan, J., \& Deharveng, L. 1986, \aap, 155, 297

\reference{CHE74} Chevalier, R.A. 1974, \apj, 188, 501

\reference{CHU90} Chu, Y.--H., \& Mac~Low, M.--M. 1990, \apj, 365, 510

\reference{CHU94} Chu, Y.--H., \& Kennicutt, R.C. 1994, \apj, 425, 720

\reference{CHU95} Chu, Y.--H., Chang, H.--W., Su, Y.--L., \& Mac~Low, 
M.--M. 
1995, \apj, 450, 156 

\reference{COX74} Cox, D.P., \& Smith, B.W. 1974, \apjl, 189, L105

\reference{DRI93} Drissen, L., Roy, J.--R., \& Moffat, A. F. J. 1993, 
\aj, 
106, 1460 

\reference{DUR95} Duric, N., Gordon, M.S., Goss, W.M., Viallefond, F., \& 
Lacey, C. 1995, \apj, 445, 173

\reference{ELM94} Elmegreen, B.G. 1994, \apj, 427, 384

\reference{FAB87} Fabbiano, G., \& Trinchieri, G. 1987, \apj, 315, 46

\reference{HAR97} Hartmann D., \& Burton W.B. 1997, in An Atlas of 
Galactic 
Neutral Hydrogen, Cambridge University Press

\reference{KIM98} Kim, S., Chu, Y.--H., Staveley--Smith, L. \& Smith, 
R.C. 1998, \aj, submitted

\reference{LAC97} Lacey, C., Duric, N., \& Goss, W.M. 1997, \apjs, 109,
417

\reference{MAC98}  Mac~Low, M.--M., Chang, T.H., Chu, Y.--H., \& Points, 
S.D. 1998, \apj, 493, 260

\reference{MAR94} Martimbeau, N., Carignan, C., \& Roy, J.--R. 1994, \aj, 
427, 656

\reference{MIL95} Miller, B.W. 1995, \apj, 446, L75

\reference{MIL94} Miller, B. W., \& Hodge, P. 1994, \apj, 427, 656

\reference{MUE76} Mueller, M. W., \& Arnett, W.D. 1976, \apj, 210, 670

\reference{PUC92} Puche, D., Westpfahl, D., Brinks, E., \& Roy, J.--R. 
1992, \aj, 103, 1841

\reference{RAY77} Raymond, J.C., \& Smith, B.W. 1977, \apjs, 35, 419

\reference{ROS84} Rosa, M., Joubert, M., \& Benvenuti, P. 1984, \aaps, 
57, 361

\reference{SNO94} Snowden S.L., McCammon D., Burrows D.N., \& Mendenhall 
J.A. 
1994, \apj, 424, 714

\reference{TEN88} Tenorio--Tagle, G., \& Bodenheimer, P. 1988, \araa, 26,
145

\reference{WEA77} Weaver, R., McCray, R., Castor, J., Shapiro, P., \& 
Moore, 
R. 1977, \apj, 218, 377 

\reference{WAL98} Walter, F., \& Brinks, E., 1998, \aj, submitted (WB98)
\end{references}
\end{document}